\documentclass[prl, twocolumn, floatfix]{revtex4}
\usepackage{graphicx}
\usepackage{amsmath, amsfonts, amssymb, bm}
\usepackage{braket}
\usepackage{color}
\usepackage[utf8]{inputenc}
\usepackage{mathrsfs}
\usepackage{amsbsy}

\begin{document} 

\title{ Probing atoms by periodically modulated electron bunches } 

\author{ A. B. Voitkiv$^1$, E. Schneidmiller$^2$ and T. Pfeifer$^3$}  
\affiliation{ $^1$ Institute for Theoretical Physics I, Heinrich-Heine-University D\"usseldorf, 
Universit\"atsstrasse 1, D-40225 D\"usseldorf, Germany \\ 
$^2$ Deutsches Elektronen-Synchrotron DESY, Notkestr. 85, 22607 Hamburg, Germany \\  
$^3$ Max-Planck-Institut for Nuclear Physics, Saupfercheckweg 1, 69117 Heidelberg, Germany
} 

\date{\today} 

\begin{abstract} 
When passing through an undulator in a Free Electron Laser, 
dense bunches of relativistic electrons split into micro-bunches, 
attaining a periodic space-time structure.  
We show that the field of such periodically modulated bunches 
is tremendously influenced by 
coherence effects, resulting in a novel type of beam-atom interaction.  
Our results indicate that employing such bunches 
(alone or in combinations with the radiation they emit) 
offers a multitude of new opportunities for exploring 
atomic dynamics on a femtosecond time scale. 
\end{abstract} 

\pacs{34.10+x, 34.50.Fa} 

\maketitle 

Interactions of photons and charged particles 
with matter belong to the basic processes studied by 
physics and results of these studies have numerous important 
applications in other fields of research,  
ranging from astrophysics to biology and chemistry.    

Until recently beams of charged particles (electrons, ions) had relatively low densities 
and their collisions with atomic systems were driven by interactions  
with individual particles of the beam \cite{Weigold}-\cite{abv_review}.  
However, nowadays high-density bunches of extreme 
relativistic electrons are generated by conventional 
and laser or plasma wakefield 
accelerators \cite{linac} - \cite{PWFA2}.  
When colliding with an atomic target 
a significant fraction of bunch electrons may 
hit the target within its transition time.  
As a result, these electrons could coherently interact   
with the target, that strongly affects the target dynamics \cite{we-2025}.  

In a Free Electron Laser (FEL) a dense bunch of highly relativistic electrons from 
a linear accelerator passes through an undulator, generating nearly monochromatic 
intense radiation \cite{FEL1}-\cite{fel-book}.     
When passing through an undulator the bunch  -- 
due to the interaction with the magnetic field of the undulator and 
the emitted radiation -- splits into micro-bunches, attaining a periodic space-time structure 
of a 'diffraction grating' which moves with an extreme relativistic velocity. 
According to our estimates, depending on the bunch and undulator parameters, 
the 'diffraction grating' may keep its shape over distances 
$ \sim 1 - 10 $ m after leaving the undulator. 

In this communication we show that, due to exceptional properties of such 'grating'-projectiles 
-- a periodic structure, a large number (up to $ \sim 10^6 $ - $ 10^7 $) of electrons 
in a micro-bunch and a very short ($\lesssim 100 $ as) micro-bunch   
duration, their collisions with atoms are driven by 
a novel type of a beam-atom interaction.  
Its novelty originates in a multi-level coherence, 
where a coherent action of the micro-bunches re-shapes the high-frequency  
part of the bunch field into nearly monochromatic lines, whose intensity 
is strongly amplified by a coherent action of electrons within each micro-bunch, 
and a coherent action of the entire bunch  
creates an extreme intense  
maximum at low frequencies. Exploiting this interaction   
offers new opportunities for exploring atomic dynamics on very short time scales. 

The field of a highly relativistic charged particle 
can be though of as consisting of the so called 'equivalent photons' \cite{equiv-ph}-\cite{jack}. 
We will show that the high-frequency part of the bunch-atom interaction is 
transmitted by coherent equivalent (CE) photons   
whose energies coincide with energies of the real photons 
emitted by the bunch in the undulator and whose intensities 
can be made close to those of the real photons.

Let the target atom, which is at rest in the laboratory frame, 
be bombarded by a bunch of extreme relativistic electrons. 
The bunch electrons move in the laboratory frame with a velocity $\bm v$ along 
the $z$-axis, $\bm v = (0,0,v)$. We shall suppose that the bunch 
was modulated along this axis that transformed it into 
a train of equidistantly separated 
micro-bunches with a similar shape.  

The cross section  $ \frac{ d \sigma_{\rm fi}}{ d^2 \bm q_\perp } $  
for atomic transition, differential in the transverse momentum transfer  
$\bm q_{\perp}$ ($ \bm q_{\perp} \cdot \bm v = 0 $) to the atom,  
is given by    
\begin{eqnarray} 
\frac{ d \sigma_{\rm fi}}{ d^2 \bm q_\perp }  & = & 
\frac{ d \sigma^{\rm (1 \, el)}_{\rm fi}}{ d^2 \bm q_\perp }    
\times \Big \{ \vert F(\bm q) \vert^2  +  N_t \Big \}.         
\label{v1}
\end{eqnarray} 
Here, $\frac{ d \sigma^{\rm (1 \, el)}_{\rm fi}}{ d^2 \bm q_\perp }$  is the cross section in the collision 
with a single electron from the bunch. It reads   
\begin{eqnarray} 
\frac{ d \sigma^{\rm(1 \, el)}_{\rm fi}}{ d^2 \bm q_\perp } \! = \! 
\frac{ 4 \, e_0^4 }{ \hbar^2 v^2 } \frac{ \big \vert \langle \psi_{\rm f} |  
e^{i \bm q \cdot \bm r} \!  
- \!\! \frac{ v }{ 2 m_e c^2 } \! \big( \! e^{i \bm q \cdot \bm r} \hat{p}_z \! + \! 
\hat{p}_z e^{i \bm q \cdot \bm r} \big)  
\vert \psi_{\rm i} \rangle \big \vert^2 }  
{\left( \bm q^2 - \frac{ \omega_{\rm fi}^2}{c^2}   
\right)^2 },      
\label{em2} 
\end{eqnarray}
where $e_0$ and $m_e$ are the electron charge and mass, respectively, 
$\bm r = (x,y,z)$ is the coordinate of the atomic electron with respect 
to the atomic nucleus,  $\hat{p}_z = \frac{ \hbar }{ i } \frac{\partial }{ \partial z } $,
$\psi_{\rm i}$ and $\varepsilon_{\rm i}$ 
($\psi_{\rm f}$ and $\varepsilon_{\rm f}$) 
are the initial (final) atomic state and its energy, respectively,   
$\omega_{\rm fi} = (\varepsilon_{\rm f} - \varepsilon_{\rm i})/\hbar$ is the atomic transition frequency, 
$ \hbar \bm q = (\hbar \bm q_\perp, \hbar q_\parallel) $ is the momentum transfer to the atom, 
$\hbar q_\parallel = \hbar \omega_{\rm fi}/v $ is the minimum momentum transfer 
and $c$ is the speed of light.  

Further, $ N_t $ is the total number of electrons in the bunch 
and $ F(\bm q) $ is the elastic form-factor of the bunch,  
\begin{eqnarray} 
F(\bm q)  & = &  
\int \!\!d^3 {\bm R } \, \, 
\rho( \bm R) \, \,  
e^{ - i \bm q \cdot \bm R },        
\label{v3}
\end{eqnarray}
where $ \rho( \bm R) $ is 
the bunch density and the integration runs over the 
bunch volume \cite{fixed-time}.       
In (\ref{v1}) the terms $ \sim N_t $ and $ \sim \vert F(\bm q) \vert^2 $ 
correspond to the independent (incoherent) and coherent, respectively, action 
of the bunch electrons in the collision. 
The coherent channel may become very efficient 
when there is a large number of electrons in 
the coherence volume $ V_c \simeq \frac{ \pi }{ q_\perp^2 q_\parallel } $ \cite{we-2025}.

If the bunch is a train of equidistantly separated 
micro-bunches with a similar shape, we obtain 
\begin{eqnarray} 
F(\bm q)  & = & \mathcal{F}(\bm q)  \, \,  \mathcal{G}(\bm q),           
\label{v4}
\end{eqnarray}   
where   
\begin{eqnarray} 
\mathcal{F}(\bm q) = \int \!\!d^3 {\bm R } \, \, 
\rho_0( \bm R) \, \,  
e^{ - i \bm q \cdot \bm R }           
\label{v5}
\end{eqnarray}
is the elastic form-factor of a single micro-bunch 
with the density $ \rho_0( \bm R) $ and    
\begin{eqnarray} 
\mathcal{G}(\bm q) = \sum_{ k } f_{k,0} \, e^{ - i \, k \, q_\parallel \, \mathcal{L}}                      
\label{v6}
\end{eqnarray} 
is the structure factor of the bunch. In (\ref{v6}) 
the sum runs over the micro-bunches, $\mathcal{L}$ 
is the distance between them and   
$f_{k,0}$ are coefficients  
describing the relative intensities of the micro-bunches. 
This factor arises due to the interference 
between the contributions to the atomic transition 
amplitude from the interaction with different micro-bunches. 
Its form closely resembles that of the factor describing 
a multi-slit interference of light scattered off a diffraction grating.  
In our case the effective 'grating' arises due to a periodic structure 
of the electron bunch \cite{rest-frame}. 

The factor $ \vert \mathcal{G} \vert^2 $ reaches its maximum 
at $ q_\parallel \, \mathcal{L} = 2 \pi n$, where 
$n = 1, 2, 3, ...$ which corresponds to resonant frequencies,   
$ \omega_{n} = \frac{2 \pi }{ \mathcal{T} } \, n$, where 
$ \mathcal{T}  = \mathcal{ L }/v $ 
is the time interval between the arrival of two consecutive micro-bunches. 
When the number of the  micro-bunches is very large and 
$f_{k,0}$ varies slowly with $k$, $ \vert \mathcal{G} \vert^2 $ strongly peaks at 
the resonant frequencies. 

\vspace{0.25cm}  

In what follows we shall consider electron bunches by relying on 
parameters of two high-repetition-rate superconducting FEL facilities, 
FLASH \cite{FLASH} and the European XFEL \cite{EuXFEL}. 
The former, which is driven by bunches with electron energies $ 0.5$ - $1.35$ GeV,  operates in the XUV and soft x-ray regimes and comprises two undulator lines. 
One of these has been upgraded to a high-repetition-rate seeded FEL line \cite{high-rep} 
and is expected to deliver fully coherent photon pulses for user operation in the near future. 
For simplicity, we consider a special regime
proposed to improve seeding performance, 
namely two-bunch seeding \cite{two-bunch-seed}, 
in which only the fundamental harmonic 
of the FEL undulator is imprinted 
on the electron bunch \cite{many-harmon}.

Let us consider the ionization of H(1s) atoms by a (FLASH) electron bunch  
with an electron energy of $1.35$ GeV,  
a total charge of 250
pC (corresponding to $ N_t = 1.6 \times 10^9$  electrons) and a 
bunch length of $L = 30 $ $\mu$m. The bunch passed through 
a linearly polarized undulator with the period $ \lambda_U = 3.5 $ cm 
and deflection parameter $ \mathcal{ K } = 3.8$. 
With an average beta function of $7$ m in the undulator 
and the normalized emittance \cite{linac}  
$0.5$ mm$ \cdot $mrad,  
the bunch radius is $ a_0 = \sqrt{2} \sigma_x = 51$ $\mu$m. 
In the undulator   
the electron bunch splits into micro-bunches  
separated along the bunch propagation by the fundamental wavelength 
of the emitted radiation \cite{FEL1}. 
As our estimates show, 
after leaving the undulator the bunch keeps this periodic structure 
over distances of several meters. 

We shall assume that the shape of the electron micro-bunches is Gaussian,   
\begin{eqnarray} 
\rho_{k}( \bm R) = \frac{ \mathcal{N}_k }{ \pi^{3/2} \, \, a_0^2 \, l_0   } \,  
\, e^{- Z^2/l_0^2 }  \, \, e^{ - R_\perp^2/a_0^2},   
\label{v7}
\end{eqnarray} 
where $ \mathcal{N}_k $ is the number of electrons in the $k$-th micro-bunch, and 
$ l_0 $ and $ a_0 $ are the effective length and radius, respectively, 
of the micro-bunches \cite{rms}. The coefficients $ f_{k,0} $ in (\ref{v5})  
are chosen to reflect the Gaussian longitudinal profile of the entire bunch. 

The chosen bunch parameters result in the value of  
$ \lambda_1 \approx 2 \times 10^{-6} $   cm for  
the fundamental wavelength, given by  
$ \lambda_1 = \frac{ \lambda_U }{ 2 \gamma^2 } (1 + 0.5 \, \mathcal{ K }^2) $ 
\cite{FEL1}, 
for the electromagnetic radiation emitted by 
the bunch along its propagation direction. 
The fundamental frequency $\omega_1 = \frac{2 \pi \, v}{ \lambda_1 } $  
corresponds to the energy $ \hbar \, \omega_1 \approx 60 $ eV.   
We further obtain that the effective number  
of the micro-bunches $ \simeq L/\lambda_1  = 2 \times 10^3 $ and   
the number $\mathcal{N}_k$ of electrons in the most intense micro-bunches is  
$ \simeq 7 \times 10^5 $. The length $l_0$ of a micro-bunch is proportional to 
$ \lambda_1 $: $l_0 = \eta \, \lambda_1 $. We take $\eta = 0.25$, which yields   
$ l_0 \simeq  5 \times 10^{-7}$ cm  
(the corresponding micro-bunch duration is $ \approx 17 $ as). 

\begin{figure}[t] 
\vspace{-0.5cm}
\begin{center}
\includegraphics[width=0.40\textwidth]{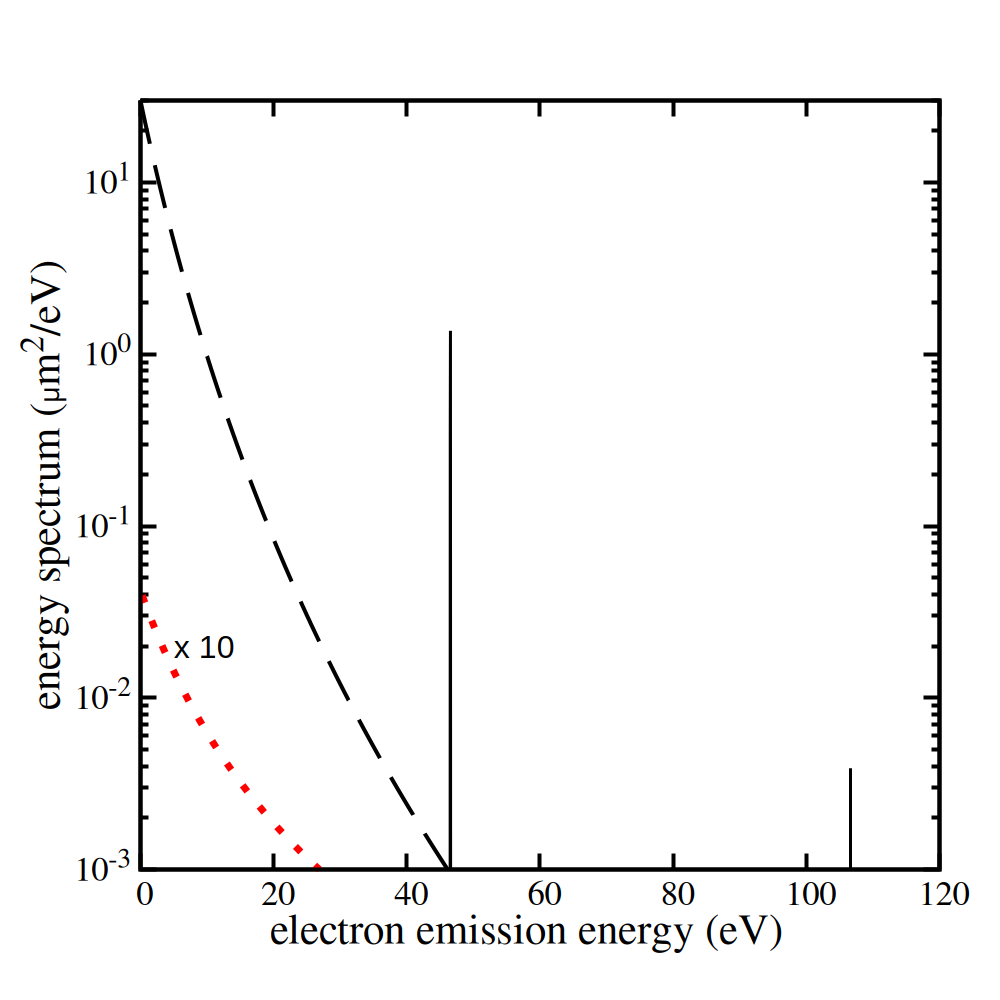}
\end{center}
\vspace{-0.95cm}
\caption{ Energy spectrum of electrons emitted in ionization of H(1s) atoms  
by a bunch of $1.35$ GeV electrons passed through an undulator. 
$N_t = 1.6 \times 10^{9}$, $ L = 30 $ $\mu$m, $a_0 = 50 $ $\mu$m, 
$ \lambda_U = 3.5 $ cm, $\mathcal{ K } = 3.8 $ and 
$l_0 = 0.005 $ $\mu$m.    
Dotted curve: all bunch electrons act incoherently (multipled by $10$).
Dashed curve: coherence within a micro-bunch is taken into account 
but the micro-bunches act incoherently. 
Solid curve: coherences within a single micro-bunch and of different micro-bunches 
are taken into account. The cross sections are given {\bf per bunch}.     
} 
\vspace{-0.1cm}
\label{figure1}
\end{figure}
   
In figure \ref{figure1} we present results of three different calculations 
for the energy spectrum of the emitted electrons.  
 
i) The dotted curve was 
obtained by assuming that all 
$1.6 \times 10^{9}$ bunch electrons interact with the hydrogen atoms 
individually (in order to make this spectrum more visible, 
in the figure it was multiplied by $10$). 

ii) The dashed curve accounts for  
the coherent action of the electrons within each 
micro-bunch but assumes that the  
micro-bunches act incoherently.       

iii) The solid curve was obtained  
by taking into account the coherent action not only   
within each micro-bunch but also  
across all micro-bunches.  

Figure \ref{figure1} demonstrates that the coherence within a micro-bunch  
strongly enhances the ionization process whereas the coherent action of the 
micro-bunches drastically changes the shape of the emission spectrum by 
compressing it into sharp peaks separated by 
the energy $ 2 \pi \hbar \, \frac{ v }{ \lambda_1 } $. 
This points to the presence of strong (quasi-) monochromatic lines 
$ \omega_n = \frac{2 \pi \, v}{ \lambda_1 } \, n$ ($n = 1,2,3,...$) 
in the energy spectrum of the CE photons 
carried by the electron bunch passed through the undulator.  

\begin{figure}[t] 
\vspace{-0.25cm}
\begin{center}
\includegraphics[width=0.40\textwidth]{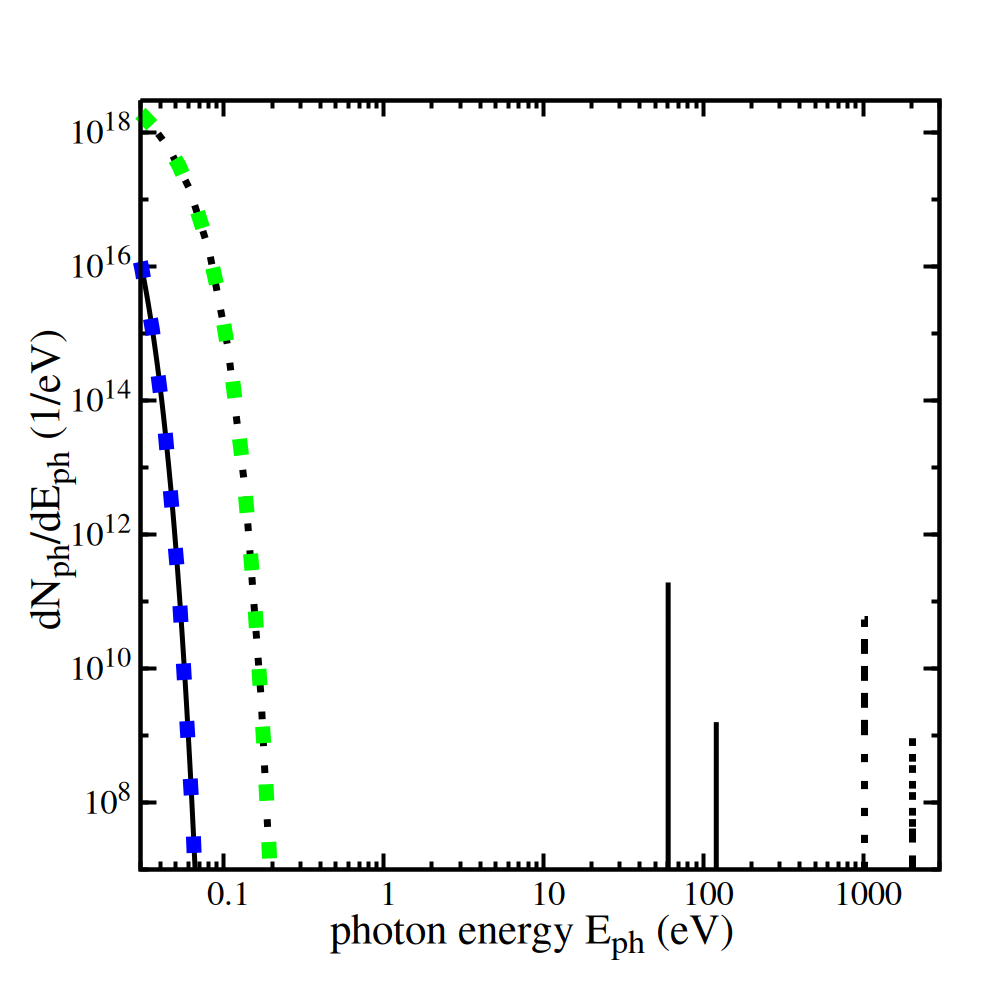}
\end{center}
\vspace{-0.95cm}
\caption{ The energy spectrum of equivalent photons 
carried by $1.35$ GeV (solid curve) and $ 14 $ GeV (dash curve) electron bunches 
(for each bunch only the first two high-frequency lines are shown). The spectrum  
of equivalent photons, carried by the corresponding unmodulated bunches,   
are shown by squares. 
For more explanations see text. 
} 
\vspace{-0.1cm}
\label{figure2}
\end{figure}  

In figure \ref{figure2} we show the energy spectrum 
of equivalent photons for two electron bunches. One of them is  
the bunch considered above. The other can be operated at  
the European XFEL, which produces coherent x-ray radiation at three undulator lines driven
by a high-energy electron bunch with energies up to $16.3$ GeV. The line SASE3 operates 
in the soft-x-ray regime, enabled by its large undulator period of $\lambda_U = 6.8$ cm 
and high deflection parameter. We consider an electron energy 
of $14$ GeV and $\mathcal{K} = 7.2$, 
resulting in a photon energy of $ \approx 1 $ keV. The electron bunch parameters 
are the same as in the FLASH example, except for a shorter length of $ 10 $ $\mu$m 
and a smaller radius $ a_0 = 24 $ $\mu$m (assuming 
an average beta function of $16$ m in the undulator). 
The considered undulator line operates in 
the self-amplified spontaneous emission (SASE) regime, which provides only partial longitudinal
coherence \cite{fel-book}. 
Full longitudinal coherence can be achieved using a self-seeding scheme 
\cite{feldhaus}, \cite{ratner}. 
Here, for illustration, we consider the fully coherent case \cite{f-SASE}.  

Two main features of the photon spectrum  
are seen in figure  
\ref{figure2}: (i) narrow high-frequency lines of 
CE photons  
(for each bunch only the first two are shown), 
whose positions practically coincide with those 
of the real photons emitted in the undulator (not shown), and 
(ii) an extremely intense maximum at very low energies,  
$ \hbar \omega_{\rm ph} \lesssim \hbar v/L $.  

\begin{figure}[t] 
\vspace{-0.5cm}
\begin{center}                                       
\includegraphics[width=0.40\textwidth]{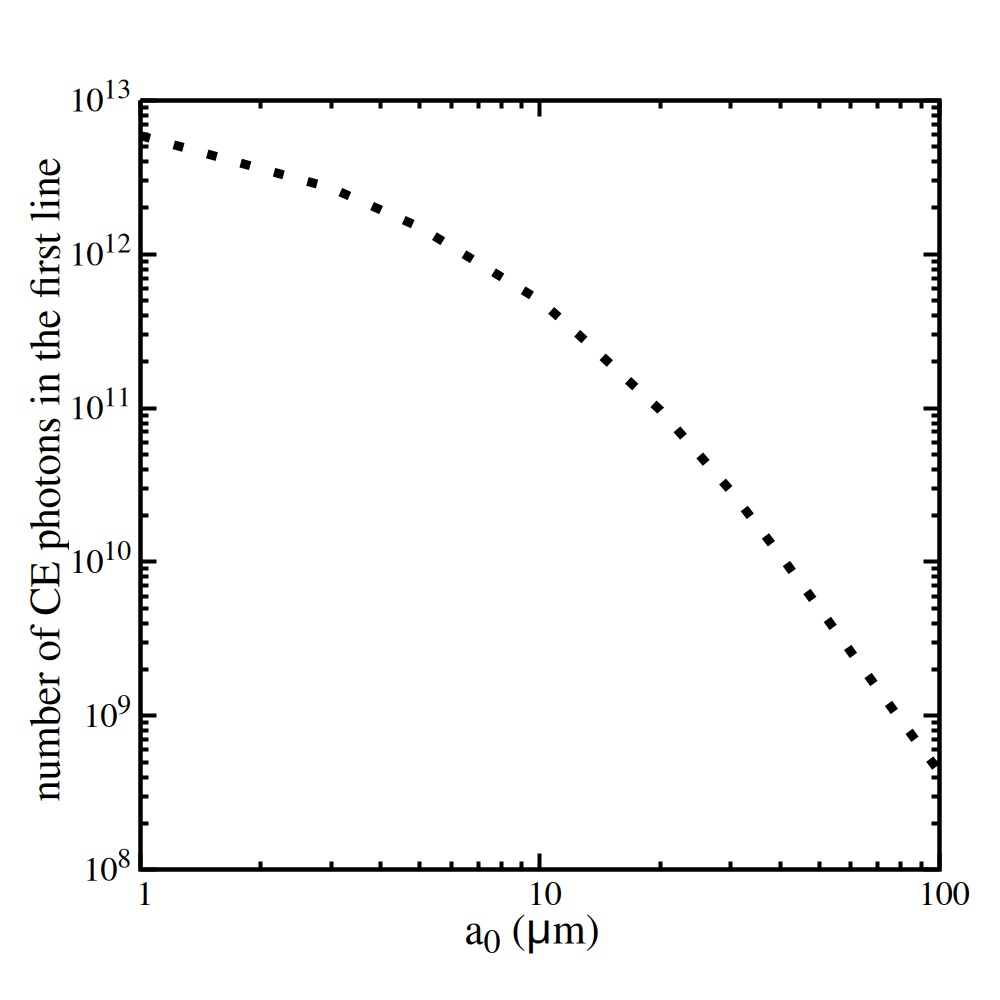} 
\end{center}
\vspace{-0.95cm}
\caption{ The number of CE photons in the line centered at 
$\hbar \omega_1 \approx 60$ eV 
as a function of the bunch radius $a_0$. 
The other parameters are $N_t = 1.6 \times 10^9$,  $1.35$ GeV, 
$ L = 30 $ $\mu$m, $ \lambda_U = 3.5 $ cm, $\mathcal{ K } = 3.8 $ and 
$l_0 = 0.005 $ $\mu$m.   
} 
\vspace{-0.1cm}
\label{figure3}
\end{figure}

The analysis shows that the integral intensity 
of each high-frequency line is inversely 
proportional to the length $L$ of 
the electron bunch and that the dependence on 
the bunch radius $a_0$ is more complicated,  
varying from $ \sim a_0^{-4} $ at $ a_0  \gg v \gamma/ \omega_n  $ 
to $ \sim \ln(1/a_0) $ at $ a_0  \ll v \gamma/ \omega_n  $. The 
$a_0$-dependence is illustrated in figure \ref{figure3}. 

\vspace{0.15cm} 

The maximum at very low frequencies 
contains equivalent 
photons with wavelengths  
comparable or exceeding the bunch length $L$ and, therefore,   
its shape \cite{1bunch} is not sensitive to micro-bunching 
(see figure \ref{figure2}).   
This maximum is so extreme intense because  
the entire bunch coherently contributes to it, 
making its intensity proportional to $N_t^2$.

Our calculations for the photon spectrum differential in 
the distance $b$ from the bunch axis show that 
even very large $b$ ($b > 100 \, a_0$, \cite{vacuum-chamber}) noticeably contribute 
to the low-frequency maximum. Therefore, its enormous intensity 
does not necessarily imply 
strong fields because the low-frequency photons 
are spread over a large volume. 

Single-pulse (unmodulated) electron bunches, whose duration greatly exceeds 
typical atomic times 
($ \tau_A \simeq 24 $ as), 
can ionize atoms via   
individual-electron---atom collisions and tunneling.    
Tunneling is a coherent mechanism 
but differs qualitatively from coherent impact ionization \cite{we-2025},  
which is efficient for short ($ \lesssim  10 \, \tau_A$) 
single-pulse bunches, proceeding    
via absorption of equivalent photons whose energy exceed 
the atomic binding energy. 

As our calculations show, 
for the bunch parameters in figures \ref{figure1} 
the tunneling ionization of H(1s) is negligible.  
However, since the tunneling ionization rate,   
$\sim \frac{F_a}{F} \exp(-2 F_a / 3 F) $ \cite{LL-III}
($F$ is the low-frequency bunch field and $F_a \sim 10^9$ V/cm 
is the characteristic atomic field), strongly depends on $F$, 
this mechanism is extremely sensitive to 
the bunch parameters \cite{we-2025}. 
We also note that long before the bunch field reaches values, 
where the tunneling becomes possible, 
it can already efficiently mix closely lying bound atomic states. 

\begin{figure}[t] 
\vspace{-0.5cm}
\begin{center}                                      
\includegraphics[width=0.40\textwidth]{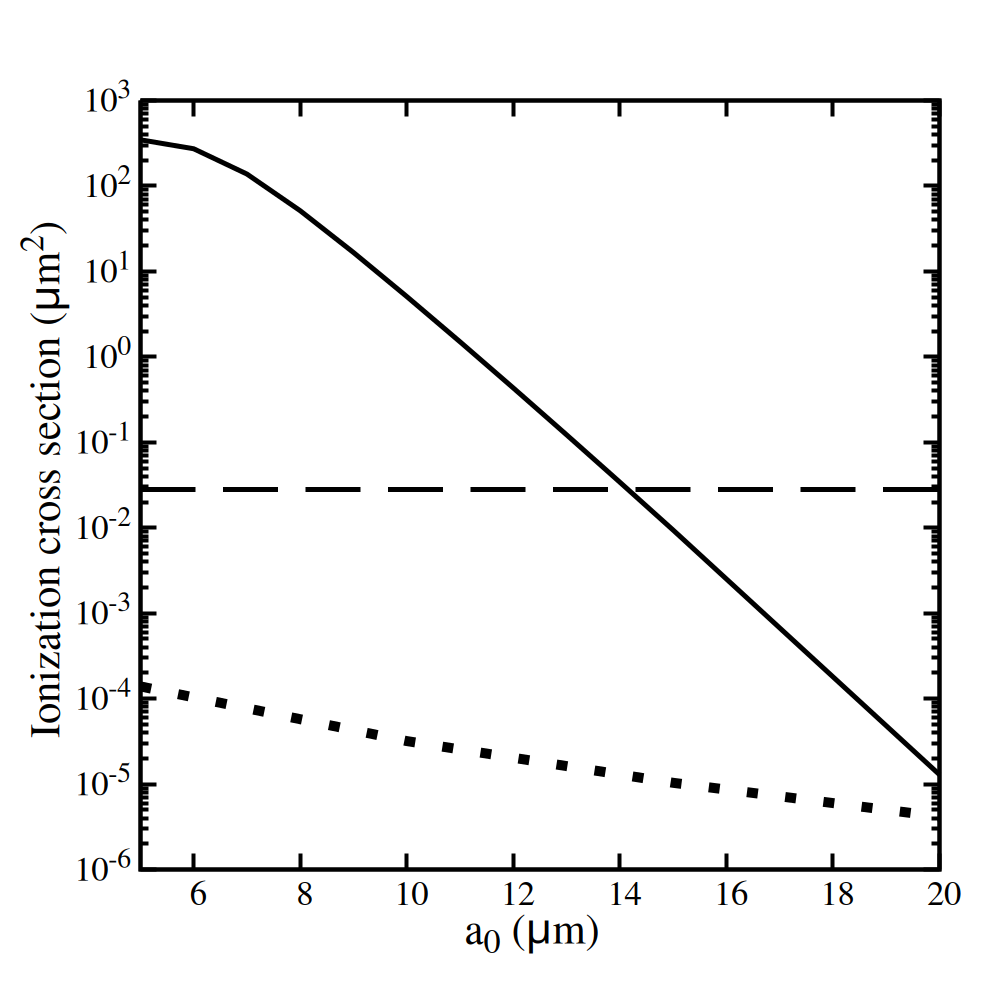} 
\end{center}
\vspace{-0.95cm}
\caption{ Ionization of H(1s) by $14$ GeV electron bunch as a function of 
the bunch radius. $N_t = 1.6 \times 10^9$,  
$ L = 10 $ $\mu$m, $ \lambda_U = 6.8 $ cm, $\mathcal{ K } = 7.2 $.  
Solid curve: tunneling ionization. Dashed curve: 
ionization by $N_t$ individual electrons. Dotted curve: ionization 
via absorption of a CE photon from 
the first high-frequency line.  
} 
\vspace{-0.1cm}
\label{figure4}
\end{figure}

The contributions of tunneling \cite{tun1}-\cite{tun2}, 
absorption of high-frequency CE photons 
and collisions with individual electrons to the ionization of H(1s) by the $14$ GeV  
bunch are shown in figure \ref{figure4}. There we see an enormous increase 
in the tunneling ionization with decreasing $a_0$ and 
inefficiency of ionization via absorption 
of the CE photons because  
their energies ($ \gtrsim 1 $ keV) greatly exceed the hydrogen 
binding energy ($\approx 13.6$ eV), making the absorpsion unlikely. 

\vspace{0.15cm} 

Thus, an electron bunch, 
passed through an undulator,  
carries an electromagnetic field, where 
nearly monochromatic high-frequency CE photons  
'co-exist' with a field that slowly varies in time.  
Such a -- perfectly synchronized --  field combination 
can, for instance, be used for exploring many-electron atoms 
where inner and outer electrons are accessed via 
the high-frequency photons and the low-frequency field, respectively, 
or studying non-dipole transitions 
in hydrogen-like ions via resonant absorption of 
a high-frequency CE photon and coupling of nearly 
degenerate excited states by the low-frequency field.  

Intensities of high-frequency CE photons 
can be further strongly enhanced 
by reducing the radius of the electron bunch.  
This can, for instance, be done by installing 
at the undulator exit a compact chicane with a photon-beam deflector, 
followed by a quadrupole triplet that focuses 
the electron beam by a factor of $\simeq 4$,  
corresponding to a beta function at the waist 
of $0.5$ m. The total length of the system would only be a few meters.

Our estimates show that for both electron bunches under consideration 
the radius can be reduced by $4$-$5$ times without 
spoiling their periodic structure. This increases 
the intensity of high-frequency CE photon lines by two orders of magnitude 
(for an illustration see figure \ref{figure3}).  

\vspace{0.15cm} 

The high-frequency CE photons closely resemble the field 
of the emitted radiation. However,  
important differences still hold, 
including a fundamental one: unlike for a real photon field, 
the integral $\int_{-\infty}^{+\infty} \bm E_{\perp} dt $ 
(where $ \bm E_{\perp} $ is the transverse part of the electric field)  
does not vanish for the CE photon field. Also, 
polarization properties can be quite different \cite{polariz}.  
This, in general, implies additional differences 
in the response of atomic systems 
when they are probed by real and equivalent photons 
produced by the same electron bunch.  

By separating the electron bunch from the emitted radiation 
at the undulator exit (while preserving its density modulation),  
the bunch can be used for probing atomic systems on a fs time scale.   
Several approaches for implementing such a separation 
have been considered  \cite{ES-6},  
including an achromatic bend or a compact chicane combined with an
absorber or deflector for the radiation.  

A key advantage of the electron beams over real photon beams generated at XFEL 
facilities is the ability to combine multiple spectral components 
within a single pulse. Beyond the combination of high- and low-frequency components, 
additional intermediate frequency bands can be produced through laser-based manipulation
in XFEL injectors \cite{new-1,new-2}, two-color seeding schemes \cite{new-3} 
with frequency-mixing techniques \cite{new-4}, 
or controlled shaping of the electron density distribution 
in bunch compressors \cite{FLASH}. 

Moreover, distinct spectral components can be generated in different temporal slices of the
same electron bunch with a controllable delay, enabling pump–probe measurements 
within a single shot. A related pump–probe concept employing real photons and 
a short ($\simeq 250$ as) broad-band single-pulse electron bunch was proposed 
in \cite{cesar}. However, in the present scheme all spectral components 
can be realized using equivalent photons alone. 
This unified approach simplifies the experimental layout 
and facilitates rapid re-configuration between different measurement modes. 

One can also probe atomic systems 
by various coherent combinations of CE photons with 
the emitted radiation.  
Here various scenarios are possible. 
For instance, one can shift azimuthal symmetry axes 
of beams of equivalent and real photons 
with respect to each other, producing  
a high-frequency field with non-trivial polarization. 
Beams of equivalent and real photons can
also be combined at different angles, 
enabling multiwave mixing \cite{new-5}, 
controlled interference effects, 
and the generation of complex polarization structures.

\vspace{0.25cm} 

In conclusion, in FELs, bunches of extreme relativistic electrons 
pass through an undulator, attaining a periodic space-time structure 
that may survive over considerable distances after leaving the undulator. 
We have shown that the interaction of such bunches with atoms  
is tremendously modified by multi-level coherence effects, 
where a coherent action of the micro-bunches 
re-shapes the high-frequency part of the bunch field into nearly 
monochromatic lines, a coherent action of electrons within each micro-bunch 
strongly amplifies intensities of these lines and   
a confluence of these two coherences occurs at low frequencies, 
resulting in an extreme intense  
maximum, created by a coherent action of the entire bunch. 

In FELs, electron bunches are employed for generating 
electromagnetic radiation.  
Our results indicate that such bunches can further be exploited   
(alone or in coherent combinations with the radiation they emit),  
offering  a multitude of new opportunities for exploring 
atomic dynamics on a femtosecond time scale.

\section*{Acknowledgement} 

We are grateful to C. M\"uller, G. Geloni, S. Liu, 
W. Decking and N. Golubeva for useful discussions.


\begin{thebibliography}{99}
 
\bibitem{Weigold}  I. E. McCarthy, E. Weigold, 
{\it Electron-Atom Collisions}  
(Cambridge University Press, 2009).  
 
\bibitem{joachain} Ph. G. Burke und C. J. Joachain, 
{\it Theory of Electron—Atom Collisions} 
(Springer US, 2013). 

\bibitem{Eich} 
J. Eichler, {\it Lectures on Ion-Atom collisions} 
(Elsevier, New York 2005).  

\bibitem{abv-book-2008} A. B. Voitkiv and J. Ullrich, 
{\it Relativistic Collisions of Structured Atomic Particles} 
(Springer, Berlin, 2008).  

\bibitem{abv_review}  
See also e.g. Chapters 49, 53, 57, 61, 67 and 69 in in { Springer Handbook of
Atomic, Molecular, and Optical Physics}, 2nd Edition,  
ed. by Gordon W. F. Drake (2023). 

\bibitem{linac} H. Wiedemann, 
{\it Particle Acceleration Physics}, 4th Edition 
(Springer, 2015). 
 
\bibitem{beam1}  
E. Esarey, C. B. Schroeder, and W. P. Leemans, 
Reviews of Modern Physics 
{\bf 81}, 1229 (2009).  

\bibitem{beam2} T. Tajima, X. Q. Yan, T. Ebisuzaki, 
Reviews of Modern Plasma Physics, 4:7 (2020). 

\bibitem{PWFA2} C. A. Lindstrøm, S. Corde, R. D'Arcy, S. Gessner, M. Gilljohann, M. J. Hogan, and J. Osterhoff, arxiv.2504.05558 (2025).  

\bibitem{we-2025} S. Kim, C. M\"uller and A. B. Voitkiv, 
arXiv 2508.17192 (submitted to Phys. Rev. Lett.) 

\bibitem{FEL1}   
C. Pellegrini, A. Marinelli, S. Reiche, 
Rev. Mod Phys. {\bf 88}, 015006 (2016).  

\bibitem{FEL2}    
E. Hemsing, G. Stupakov, Dao Xiang, A. Zholents, 
Rev. Mod Phys. {\bf 86}, 897 (2014).  

\bibitem{FEL3}   
S. Varró (ed), {\it Free Electron Lasers} (IntechOpen, 2012). 

\bibitem{fel-book} E. L. Saldin, E. A. Schneidmiller, and M. V. Yurkov, 
{\it The Physics of Free Electron Lasers}, (Springer-Verlag, Berlin, 2000).
 
\bibitem{equiv-ph}    
Equivalent photons, which represent the field of 
a highly relativistic charged particle moving with 
a {\it constant} velocity, 
may closely resemble real photons that forms the basis 
of the Weizs\"acker-Williams approximation widely 
used in high-energy physics (for a review see  e.g. \cite{WWA}).  
A simple and illustrative discussion of this approximation 
can be found in \cite{jack}.

\bibitem{WWA} V. M. Budnev, I. F. Ginzburg, G. V. Meledin and  V. G. Serbo, 
Phys. Rep. {\bf 15}, 181–282 (1975); 
C. A. Bertulani and G. Baur, Phys. Rep. {\bf 163}, 299 (1988). 

\bibitem{jack} J.D. Jackson, {\it Classical Electrodynamics}, 3rd ed., (Wiley,
New York, 1999).

\bibitem{fixed-time} 
The bunch density and volume are taken at an arbitrarily fixed moment of time. 

\bibitem{rest-frame} In the rest frame of the beam the atomic transitions 
are caused by inelastic scattering of an extreme relativistic atom 
on a 'diffraction grating' composed  of groups of 'static' electrons.  

\bibitem{FLASH} W. Ackermann, G. Asova, V. Ayvazyan, A. Azima, N. Baboi, J. B\"ahr, 
 V. Balandin, B. Beutner, A. Brandt, A. Bolzmann, R. Brinkmann, 
 O. I. Brovko, M. Castellano, P. Castro, L. Catani, E. Chiadroni,
S. Choroba, A. Cianchi, J. T. Costello, D. Cubaynes, et al, 
Nature Photonics {\bf 1}, 336 (2007). 

\bibitem{EuXFEL} W. Decking, S. Abeghyan, P. Abramian, A. Abramsky, A. Aguirre, C. Albrecht, P. Alou, M. Altarelli, P. Altmann, K. Amyan, V. Anashin, E. Apostolov, K. Appel, D. Auguste, V. Ayvazyan, S. Baark, F. Babies, N. Baboi, P. Bak, V. Balandin, et al, 
Nature Photonics {\bf 14}, 391 (2020).

\bibitem{high-rep} L. Schaper, S. Ackermann, E. Allaria, P. Amstutz, K. Baev, M. Beye, C. Gerth, I. Hartl, W. Hillert, K. Honkavaara, M. M. Kazemi, T. Lang, P. Niknejadi, F. Pannek, J. R\"onsch-Schulenburg, D. Samoilenko,
E. A. Schneidmiller, S. Schreiber, M. Tischer, M. Vogt, M. Yurkov, and J. Zemella, Appl. Sci. {\bf 11}, 9729 (2021).

\bibitem{two-bunch-seed} E. Schneidmiller and I. Zagorodnov, 
Phys. Rev. Accel. Beams {\bf 27}, 110703 (2024).

\bibitem{many-harmon} In the standard seeding scheme, 
many harmonics of the seed laser are present in the electron density modulation.

\bibitem{rms} The parameters $a_0$ and $l_0$ are related to 
the corresponding root-mean-square values of the beam by $ \sqrt{ \langle X^2 \rangle  } = \sqrt{ \langle Y^2 \rangle } = a_0/\sqrt{2} $ and $ \sqrt{ \langle Z^2 \rangle  } = l_0/\sqrt{2} $  


\bibitem{feldhaus}  J. Feldhaus, E. L. Saldin, J. R. Schneider, 
E. A. Schneidmiller, and M.V. Yurkov, Opt. Commun. {\bf 140}, 341 (1997).

\bibitem{ratner} D. Ratner, R. Abela, J. Amann, C. Behrens, D. Bohler, G. Bouchard, C. Bostedt, M. Boyes, K. Chow, D. Cocco, F.J. Decker, Y. Ding, C. Eckman, 
P. Emma, D. Fairley, Y. Feng, C. Field, U. Flechsig,
G. Gassner, J. Hastings, et al, Phys. Rev. Lett. {\bf 114}, 054801 (2015). 

\bibitem{f-SASE} In the SASE regime the
number of equivalent photons remains comparable, 
while the averaged spectrum is broadened by
the number of longitudinal modes. 

\bibitem{1bunch} The shape is mainly determined by 
the factor $ \exp( - \omega^2_{\rm ph} L^2/ v^2) $.

\bibitem{LL-III} L.D. Landau and E.M. Lifshitz, {\it Quantum Mechanics} 
(Pergamon, New York, 1965).     

\bibitem{vacuum-chamber} In our consideration we neglect electrodynamic environment 
but note that $b \sim 100 a_0$ would already approach the radius of vacuum chamber 
that prevents from considering larger values and strongly modifies 
the low-frequency part of the spectrum.
 
\bibitem{tun1}  Tunneling ionization is a highly nonperturbative process and, 
in order to calculate its cross section,  
we have used tunneling rate expressions from \cite{tun2},   
which apply to ionization by a constant or slowly varying electric field.

\bibitem{tun2} S. Remme, A. B. Voitkiv, G. Pretzler and C. M\"uller, 
J. Phys. {\bf B 58}, 195602 (2025);  
X. M. Tong and C. D. Lin, J. Phys. {\bf B 38}, 2593 (2005). 

\bibitem{polariz} For instance, electron bunches with azimuthal symmetry 
generate equivalent photons which are polarized along the bunch radius.

\bibitem{ES-6} E.A. Schneidmiller, Phys. Rev. Accel. Beams {\bf 25}, 010701 (2022). 

\bibitem{new-1} D. Cesar, A. Anakru, S. Carbajo, J. Duris, P. Franz, 
S. Li, N. Sudar, Z. Zhang, and A. Marinelli, Phys.
Rev. Accel. Beams {\bf 24}, 110703 (2021)

\bibitem{new-2} Ph. Amstutz, W. Helml, S. Khan, C. Mai, Ch. Gerth, Ch. Mahnke, E. A. Schneidmiller, 
arXiv:2508.14592

\bibitem{new-3} E. Ferrari, C. Spezzani, F. Fortuna, R. Delaunay, F. Vidal, 
I. Nikolov, P. Cinquegrana, B. Diviacco, D. Gauthier, G. Penco, P. R. Ribic,
E. Roussel, M. Trovo, J.-B. Moussy, T. Pincelli, L. Lounis, M.
Manfredda, E. Pedersoli, F. Capotondi, C. Svetina, N. Mahne, M. Zangrando,
L. Raimondi, A. Demidovich, L. Giannessi, G. De Ninno, M. B. 
Danailov, E. Allaria and M. Sacchi, Nat Commun {\bf 7}, 10343 (2016).

\bibitem{new-4} G. Geloni, F. Brinker, W. Decking, J. Gr¨unert, 
M. Guetg, T. Maltezopoulos, D. Noelle, S. Serkez, S.
Tomin, M. Yurkov, E. Schneidmiller, Applied Sciences 11(18):8495 (2021).

\bibitem{cesar} D. Cesar, A. Acharya, J. P. Cryan, A: Kartsev, M. F. Kling, 
A. M. Lindenberg, C. D. Pemmaraju, A. D. Poletayev, V. S. Yakovlev and 
A. Marinelli, Optica {\bf 10}, No 1, 1 (2023). 

\bibitem{new-5} F. Bencivenga, R. Cucini, F. Capotondi, A. Battistoni, R. Mincigrucci, E. Giangrisostomi, A. Gessini,
M. Manfredda, I. P. Nikolov, E. Pedersoli, E. Principi, C. Svetina, P. Parisse, F. Casolari, M. B. Danailov, M. Kiskinova and C. Masciovecchio, Nature {\bf 520}, 205 (2015). 

\bibitem{ES-7} E. A. Schneidmiller and M. V. Yurkov, Phys. Rev. ST Accel. Beams {\bf 16}, 
110702 (2013). 

\bibitem{origin} The results for the cross sections do not depend on 
the choice of the origin: it can be an arbitraty point 
which is at rest in the laboratory frame. 

\bibitem{schiff} L. I. Schiff, 
Rev. Sci. Instrum. {\bf 17}, 6 (1946). 

\end{thebibliography}
\end{document}